\documentclass[journal, onecolumn,12pt, draftclsnofoot]{IEEEtran}

\usepackage{amsmath,amssymb,amsfonts}
\usepackage{graphicx}
\usepackage{xcolor}
\usepackage{kotex}

\usepackage{caption}
\usepackage{subcaption}

\usepackage{ulem}

\begin{document}
%
\title{Towards 6G Hyper-Connectivity: Vision,\\ Challenges, and Key Enabling Technologies}
%
%

\author{Howon~Lee, 
        Byungju~Lee,
        Heecheol~Yang, 
        Junghyun~Kim, 
        Seungnyun~Kim,\\ 
        Wonjae~Shin, 
         Byonghyo~Shim
         , and H. Vincent Poor
        \thanks{H. Lee is with
        Hankyong National University. 
        B. Lee is with
        Incheon National University. 
        H. Yang is with
        Chungnam National University. 
        J. Kim is with 
         Sejong University. 
        W. Shin is with 
        Ajou University. 
        %
        S. Kim and B. Shim are with 
        Seoul National University. 
           H. V. Poor is with 
        Princeton University. 
        (Co-first authors: H. Lee, B. Lee, H. Yang, and J. Kim) (Corresponding authors: W. Shin and B. Shim)}}

%
%


\maketitle

\begin{abstract}
Technology forecasts anticipate a new era in which massive numbers of humans, machines, and things are connected to wireless networks to sense, process, act, and communicate with the surrounding environment in a real-time manner. 
To make the visions come true, the sixth generation (6G) wireless networks should be \textit{hyper-connected}, implying that there are no constraints on the data rate, coverage, and computing. In this article, we first identify the main challenges for 6G hyper-connectivity, including terabits-per-second (Tbps) data rates for immersive user experiences, zero coverage-hole networks, and pervasive computing for connected intelligence. 
To overcome these challenges, we highlight key enabling technologies for 6G such as distributed and intelligence-aware cell-free massive multi-input multi-output (MIMO) networks, boundless and fully integrated terrestrial and non-terrestrial networks, and communication-aware distributed computing for computation-intensive applications.
%
We further illustrate and discuss the hyper-connected 6G network architecture along with open issues and future research directions.
\end{abstract}


\IEEEpeerreviewmaketitle

\section{Introduction}

Moving toward the post-pandemic world, we are experiencing a new, previously inconceivable change in our lives. A large part of our daily activities are going online, and we are witnessing numerous use-cases in shopping, mobility, healthcare, and manufacturing. As a platform to embrace this dramatic change, 
the fifth generation (5G) wireless systems have been standardized and commercialized in 2019, and as of now, 5G services have been rolled out in more than 70 countries.
Notwithstanding the many technological advances brought by 5G, there are still many hurdles and challenges in realizing an envisioned boundless and hyper-connected society where intelligent things such as machines, vehicles, sensors, and robots are seamlessly connected both in physical and virtual spaces without limitation on the data rates and/or transmission delays \cite{Samsung6G}. To back up these disruptive changes and therefore make the slogan a reality, we need to prepare the sixth generation (6G) of wireless networks by identifying key enabling technologies that cannot be achieved through a simple evolution of 5G. 

One important expectation in the 6G era is that 
machines and things will be the main consumers of mobile data traffic, and thus there will be more than 55 billion devices connected to the Internet by 2025. These devices will continuously sense, process, act, and communicate with the surrounding environment, generating more than 73 zettabytes of data per year \cite{IDC}. 
Since 6G wireless networks should support a variety of form factors with different service requirements and hardware capabilities, enabling technologies should be dearly distinct from those of 5G in various aspects. 

\begin{figure*}
\centering 
\includegraphics[width=6.7in]{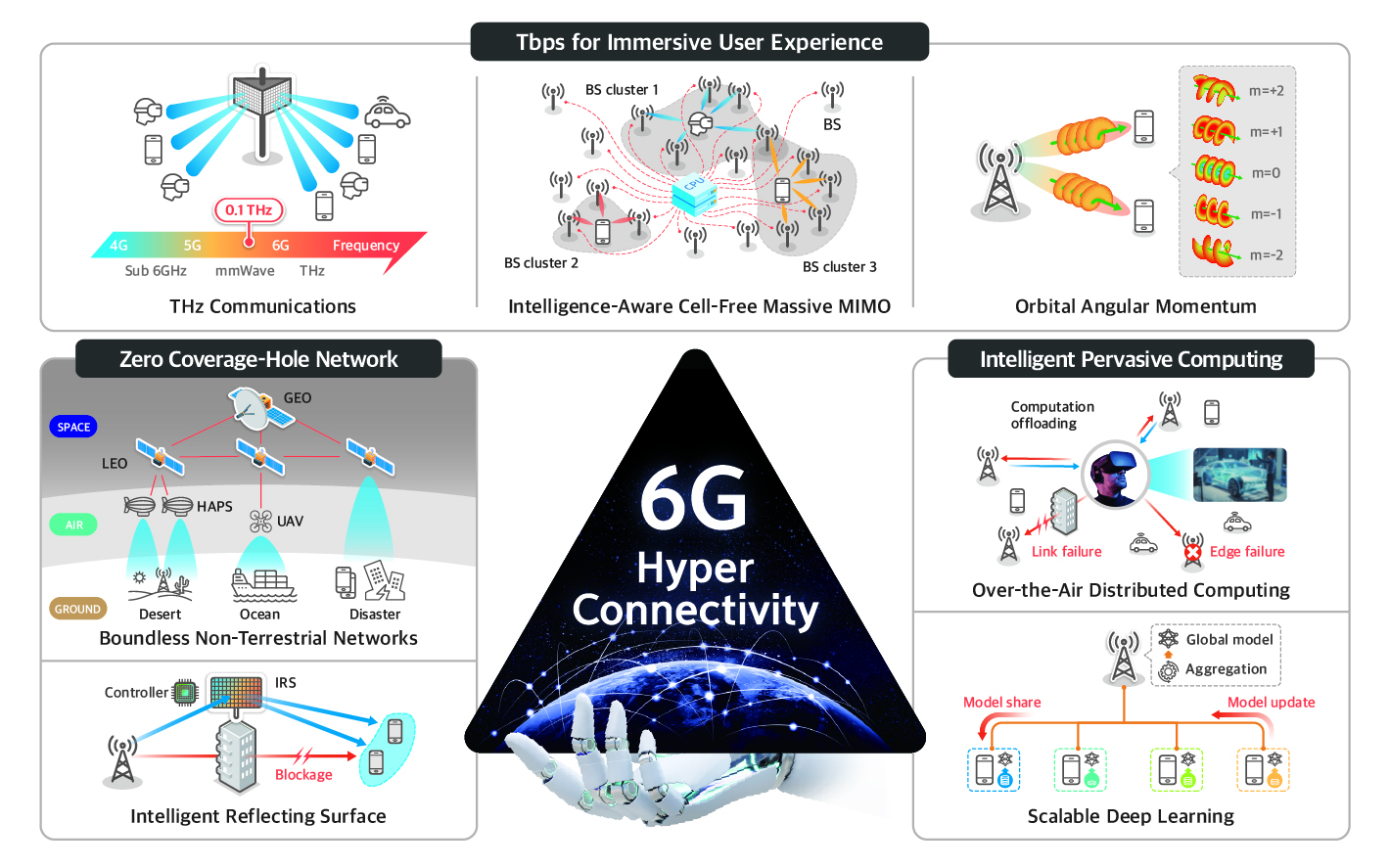} 
\caption{Key concepts of 6G hyper-connectivity.} 
\label{fig:concept2}
\end{figure*}

The primary purpose of this article is to discuss key challenges arising from these considerations. These include the provision of terabits-per-second (Tbps) wireless connectivity, zero coverage-hole networks, and pervasive computing for connected intelligence, along with enabling technologies that can be used as a core for the 6G system. The key concepts of 6G hyper-connectivity are illustrated in Fig. \ref{fig:concept2}.

After the introduction, the rest of this article is organized as follows. In Section II, we address the challenges facing 6G hyper-connectivity. In Section III, we present the key enabling technologies for realizing this 6G vision. In Section IV, we discuss future research issues and provide our concluding remarks.

\section{Challenges and Prospects Facing 6G Hyper-Connectivity} \label{sec:2}

\subsection{Tbps for Immersive User Experience}
Since the target of 5G is to support $20$ gigabits-per-second (Gbps) peak data rate and $0.1$ Gbps user-experienced data rate, it is not too difficult to support advanced multimedia services such as $8$K augmented reality (AR) streaming, $16$K virtual reality (VR) streaming, and extended reality (XR) requiring  beyond-gigabit data rates.
It is however not easy to support truly immersive mobile services such as digital twin or metaverse since the virtual model accurately reflecting the physical world using XR devices or high-fidelity mobile holographic displays requires a rate of up to Tbps \cite{Samsung6G}.
%
%
For example, a full-view $24$K VR transmission with a $64$ pixels per degree (PPD) minimum resolution, a $200\,\text{Hz}$ minimum refresh rate, and a $300\!\!:\!\!1$ compression rate requires at least $6.37$ Gbps.
Clearly, this level of rate cannot be supported anytime anywhere in the current 5G systems since the typical user-experienced data rate of 5G is in the range of 0.1$\sim$1 Gbps. 
%

One way to support emerging services requiring tremendous data rates is to exploit the terahertz (THz) frequency band. 
%
The THz band offers colossal spectrum but the THz waves suffer high attenuation due to the attendant significant path loss, molecular absorption, and atmospheric attenuation. 
One straightforward option to overcome these problems is to densify the network, meaning that we reduce the distance between the base station (BS) and mobile device. 
%
%
In this ultra-dense scenarios, pencil beamforming via ultra-massive multi-input multi-output (UM-MIMO) plasmonic nano-antenna arrays can be useful in providing 10 to 100$\times$ improvement in data rate \cite{UM-MIMO}.
%
While the densified small cells equipped with massive multi-input multi-output (MIMO) systems are a good fit for THz communication systems, their deployment will make the notions of cell and handover obsolete.
%
Thus, we need to design the cell-free networks from the scratch by considering various technical challenges, such as beam tracking/management, user association and BS coordination, and synchronization and initial access. 
%
%
Other key enabling technologies to support Tbps data rates include orbital angular momentum (OAM) multiplexing, full-duplex, and spatial modulation-MIMO (SM-MIMO).

\vspace{-2mm}
\subsection{Zero Coverage-Hole Networks}

Another more important goal of 5G systems is to provide  advanced wireless connectivity for a wide variety of vertical applications including self-driving cars and drones, robot-aided smart factories and healthcare, and remote control for disaster scenarios.
%
To accommodate such applications,
wireless access should be available anywhere, meaning that there should be no coverage holes. 
Although many network operators claim nationwide coverage for 5G, there are still many coverage holes in places such as tunnels, basements, elevators,  mountains, and oceans. 
Moreover, higher frequencies (3.5 GHz at FR1 and 28 GHz at FR2) used in 5G systems tend to be vulnerable to blockage owing to their high directivity and severe path loss. 
For these reasons, more often than not we experience link failure in our daily use of 5G services.

Recently, several new approaches to enhance the coverage and ensure universal and seamless connectivity have been proposed.
An intelligent reflecting surface (IRS), an approach that provides an artificial path in the absence of line-of-sight (LOS) links, is a viable candidate 
to eliminate coverage holes, particularly in dense urban areas \cite{RIS}.
Wave-controlled holographic MIMO surfaces that flexibly shape electromagnetic waves using sub-wavelength metallic or dielectric scattering particle structures is another good option.
In remote areas such as deserts, oceans, and mountain areas, in which the deployment of terrestrial networks (TNs) is almost impossible, non-terrestrial networks (NTNs) with satellites, high-altitude platform stations (HAPSs), and unmanned aerial vehicles (UAVs) can be an attractive solution because they ensure coverage without relying on the ground infrastructure.
%
%
By the deliberate combination of TNs and NTNs \cite{SAGINnew}, one can dramatically improve the coverage and thus generate the network with virtually no coverage hole.





   \begin{figure*}
       \begin{subfigure}{0.646\textwidth}
           \includegraphics[width=1\textwidth]{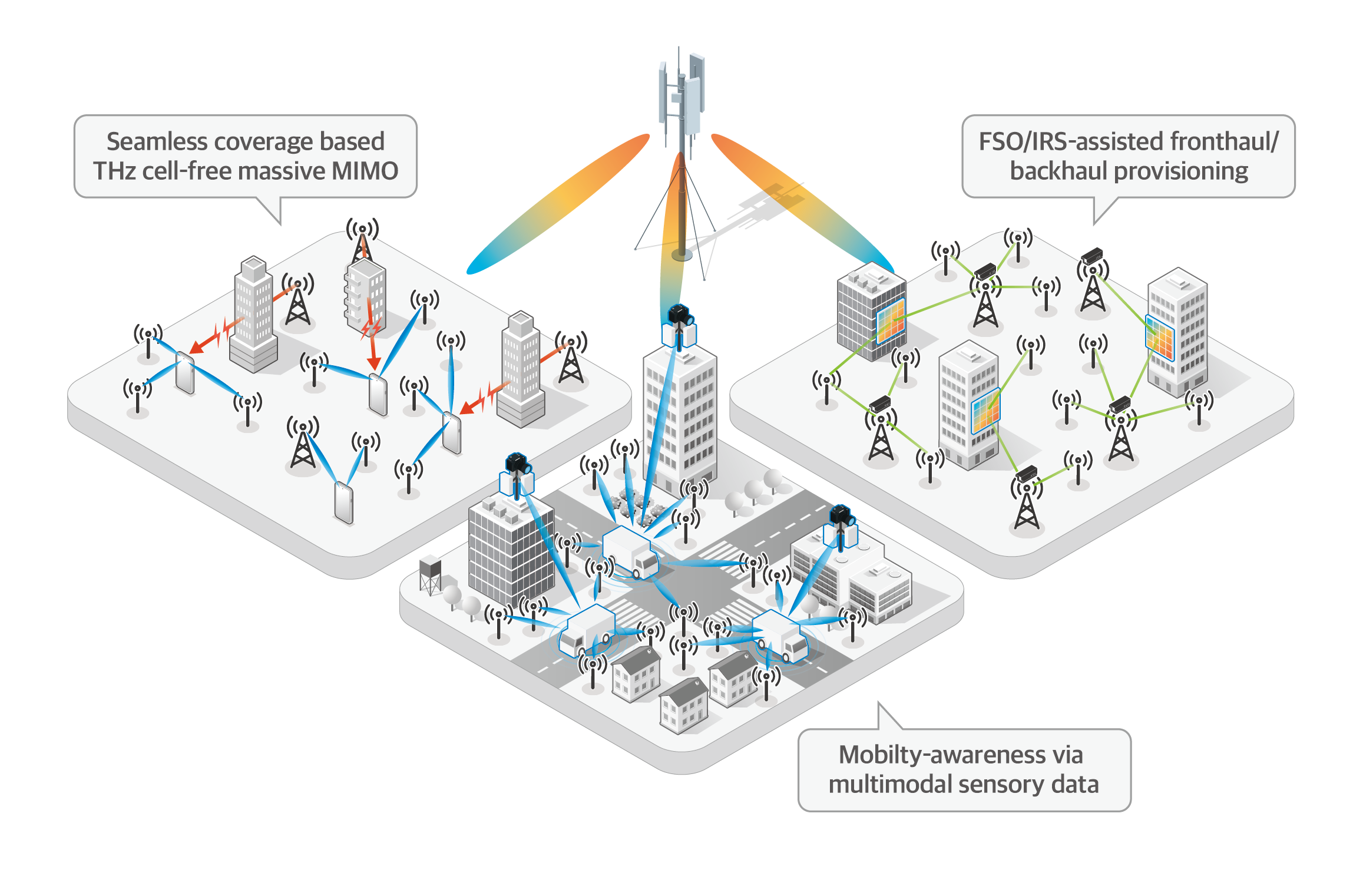} 
           \caption{}
           \label{fig:cell_free1}
       \end{subfigure}
       \begin{subfigure}{0.354\textwidth}
           \hspace{-8mm}
           \includegraphics[width=1.1\textwidth]{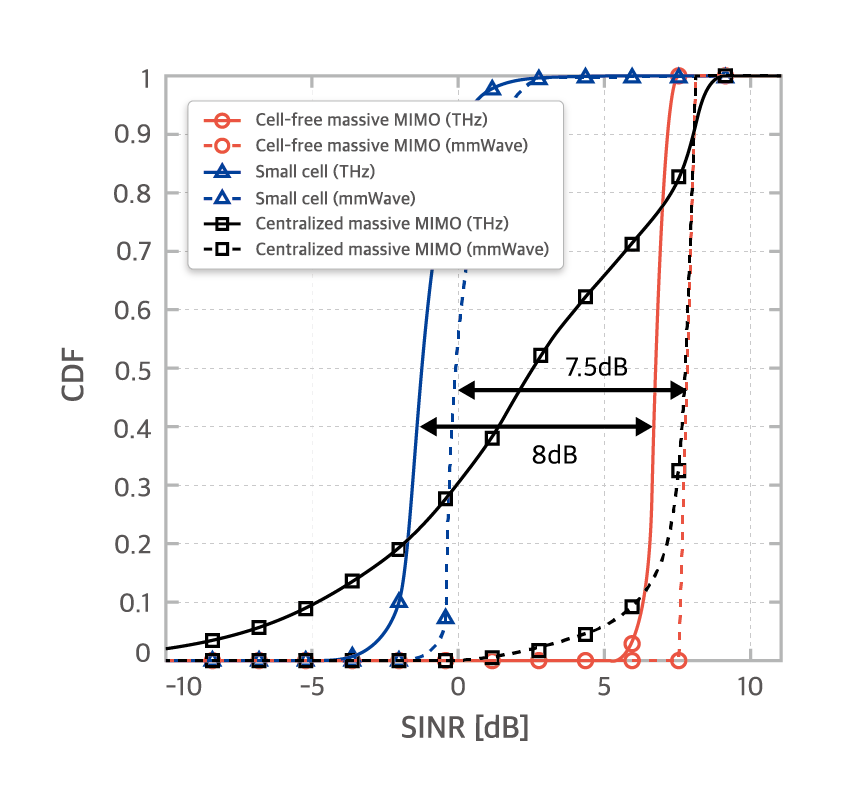} 
           \vspace{5.9mm}  
          \hspace{8mm}          \caption{}
           \label{fig:cell_free2}
       \end{subfigure}
       \caption{Left-hand side: Illustrations of 6G cell-free massive MIMO network; right-hand side: CDFs of user SINR in cell-free massive MIMO networks, small cell, and centralized massive MIMO networks.}
       \label{fig:cell_free}
   \end{figure*}
  

\subsection{Pervasive Computing for Connected Intelligence} 

With the advent of truly immersive mobile services, such as XR and metaverse, computational overhead has become a serious concern since these services require high energy consumption, end-to-end latency, and processor and memory overhead.
In 5G, mobile edge computing (MEC) has been widely used to delegate computation-intensive tasks to the edge (a local server near a BS). However, MEC would not be enough to support future applications.
In a VR service, for example, a mobile VR headset must provide photorealistic and $6$ degrees of freedom (6-DoF) rendered images within 20 ms of motion-to-render-to-photon (M2R2P) latency.
%
%
To control this overhead solely in a mobile device or offload it to a nearby BS will be by no means easy so that new paradigm to deal with the surge of computation is required.
%
%

%
Fortunately, as the density of wireless networks increases and beyond-gigabit transmissions can be supported using mmWave and THz band transmissions,
one can exploit computational resources in nearby BSs and mobile devices that are equipped with high performance computing processors.
%
%
For instance, a vehicle equipped with a high-performance battery ($1000\,\text{Wh/L}$) and powerful graphics processing units (GPUs) can perform environment sensing, learning, and computation-oriented tasks and send the results to the destination device. 
Particularly, computation offloading is useful for the artificial intelligence (AI) applications, such as object detection and classification, speech recognition, and metaverse control, since these tasks usually require heavy deep neural network (DNN) computation, but the inference results are often very simple.

In order to facilitate the boundless use of computation resources over the network, one should use advanced wireless distributed computing techniques.
Decentralized learning obtaining a global model from localized data via peer-to-peer communication among mobile devices can be one good choice.
%
When security matters, federated learning where the mobile devices jointly train the global model without exchanging the local dataset with the central server would be a proper choice.
Yet another option to support large-scale models is split learning, in which large-scale DNNs are divided into several pieces and then trained separately on mobile devices.

\section{Key Enabling Technologies for 6G Hyper-Connectivity}

In this section, we discuss key enabling technologies to achieve beyond Gbps for immersive user experience, zero coverage-hole networks, and pervasive computing for connected intelligence.


\subsection{Distributed $\&$ Intelligence-Aware Cell-Free Massive MIMO Network}



In the 6G era, a network will be heavily densified and thus, the areal density of BSs will be comparable to, or even surpass, the density of mobile devices.
As the coverage of a cell gets smaller, the distance between BS and mobile will also be shorter.
To make the most of the proximity between BSs and mobile devices, a network should be designed such that a group of small cells coherently serve users in a user-centric manner (see in Fig. \ref{fig:cell_free1}) \cite{cell-free}. 
In most urban scenarios, user-centric cell association will generate multiple LOS links between BSs and mobile so that the mobile can easily achieve a high degree of macro-diversity gain and spectral efficient improvement.
%


To evaluate the efficacy of the cell-free massive MIMO network, we plot cumulative distribution functions (CDFs) of user signal-to-interference-and-noise ratio (SINR) in mmWave ($100\,\text{GHz}$) and THz ($1\,\text{THz}$) communication scenarios.
%
One can see from Fig. \ref{fig:cell_free2} that the SINR variation of the cell-free massive MIMO network is much smaller, meaning that the cell-free massive MIMO network can well suppress the inter-cell interference through the cooperative beamforming of BSs. 
For example, in the THz systems, the cell-free massive MIMO network can provide at least $7\,$dB SINR to more than $95\%$ of users while the conventional small cell network guarantees only -$2.5\,$dB. 
We also observe that the SINR gain of the THz band is higher than that for the mmWave band, meaning that it can effectively compensate for the huge path loss of the THz systems.
To fully enjoy the benefits of cell-free massive MIMO, we need to address the following challenges.

\smallskip \noindent
\textbf{Intelligence-aware network architecture: }
In order to provide the user-centric coverage through the joint beamforming of cooperating BSs, a cell control information, channel state information (CSI), and transmit data of local BSs should be shared with the central unit (CU) through the fronthaul link. 
Sharing such a large information among all network entities will cause a significant fronthaul overhead, not to mention the tremendous computational overhead required for the channel estimation, data transmission/reception, and resource allocation \cite{cell-free2}. 
A promising approach to mitigate the overhead is the decentralized processing. One such example is the multi-agent deep reinforcement learning (MA-DRL) where the operations like beamforming, resource allocation, and user association are performed locally and individually by the DRL agent in each small cell.

\smallskip \noindent
\textbf{Multi-sensory data fusion for channel prediction and beamforming:} 
In order to realize a truly zero coverage-hole cell-free massive MIMO network, the network should have mobility-aware functions such as fast-fading channel estimation, mobile trajectory tracking, blockage prediction, and dynamic BS clustering and beamforming. 
One promising direction is to use the multi-modal sensory data (e.g., RGB depth vision information of camera, range information of LiDAR, SLAM-based user location) along with the computer vision techniques to identify the 3D shape of nearby wireless environments (e.g., obstacle locations, LOS and non-line-of-sight (NLOS) paths, and mobility patterns).
In this scheme, using a set of panoramic images obtained from the stereo camera, the DL-based object detection can extract the location (3D Cartesian coordinate $(x, y, z)$) of target objects in an image and then determine its class.
In acquiring the location information of a mobile device from input images, a convolutional neural network (CNN) that extracts the spatially correlated feature using the convolution filter can be employed.
Once the location of the mobile is identified, by measuring the angle and distance from the BS to the target object (e.g., mobile phone, vehicle, or urban air mobility (UAM)), the mmWave and THz beam can be generated without a complicated beam management process.

\smallskip \noindent
\textbf{Computation-oriented end-to-end cell-free massive MIMO network:}
In order to support the computation-intensive and/or latency-critical applications, the future 6G network should have the computation offloading functionality \cite{EdgeComp}. Computation offloading can be realized by utilizing the local computing power at the BSs as well as cloud computing at the CU. For example, latency-sensitive computation tasks will be processed at the BSs in proximity to the user, and the computation-intensive tasks can be handled via the cooperation of BSs and CU.

To ensure successful computation offloading, an integrated cell-free network design exploring both communication and computational perspectives needs to be suggested. In general, it should be guaranteed that computing operations can be delivered by reliable and low-latency communication links through a user, BSs, and CU. From a computation perspective, edge servers located at the BSs may have limited computing capabilities in terms of computing power, memory size, and energy consumption compared to the central server at the CU. In addition, since the user needs to offload computation tasks to nearby BSs, the randomness in computation latency at each BS should be taken into account to guarantee low-latency and reliable computing services.


\begin{figure*}
\centering 
\includegraphics[width=7.2in]{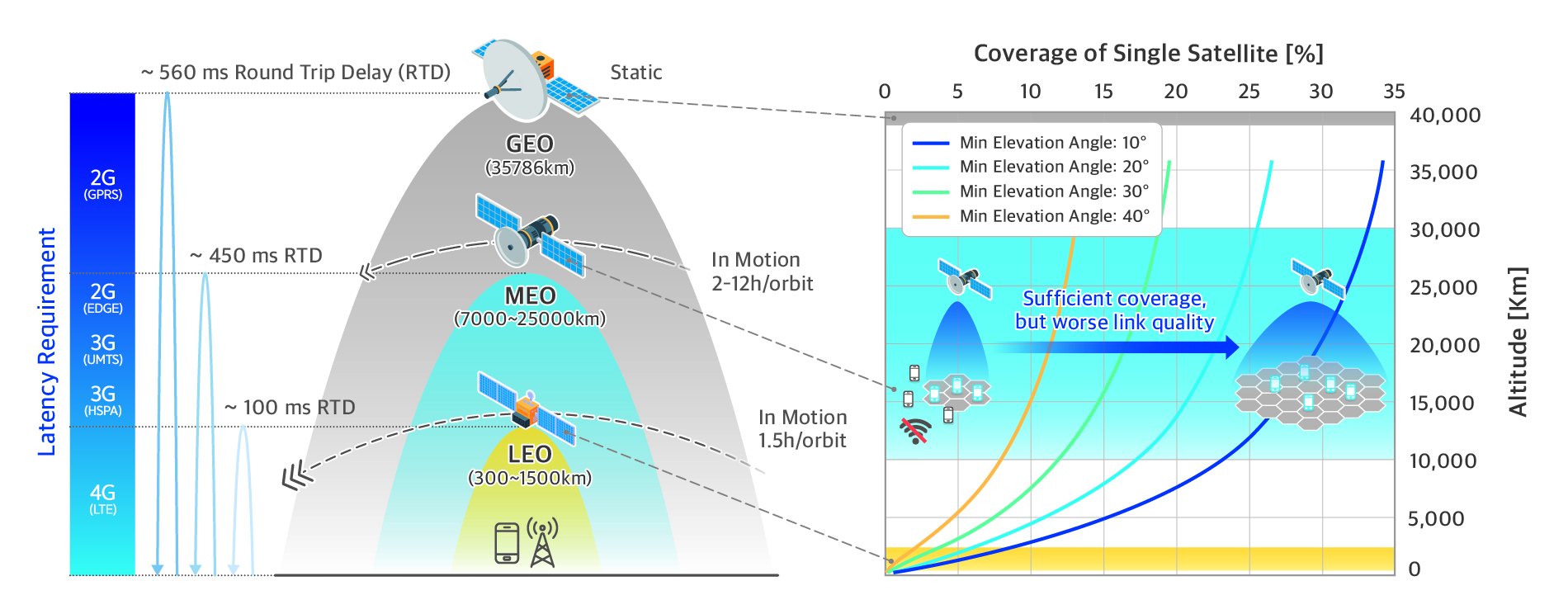} 
\caption{Satellite communication systems with different altitudes: Round-trip delay and single satellite coverage as a fraction of the Earth’s surface.} 
\label{fig:3D_connectivity} 
\end{figure*}

\smallskip \noindent
\textbf{Integrated RF and optical X-haul provisioning:}
In the cell-free massive MIMO network, fronthaul and backhaul overhead is a serious yet underrepresented problem.
Recently, advanced fronthaul/backhaul techniques have been proposed.
Replacing the current CPRI interface with the optical fiber-based fronthaul/backhaul is a simple option but will increase the CAPEX significantly.
A radio stripe network-based fronthaul and backhaul where the BSs are serially connected using daisy-chain architecture to the same cables would be a cost-effective option for the densely deployed BSs.
Also, an integrated access and backhaul (IAB) network where only a few BSs are connected to traditional fiber infrastructures while other BSs wirelessly relay the fronthaul/backhaul traffic would be a promising option. 
Yet another simple solution is free-space optical (FSO) communication where the modulated optical beams are transmitted through the free space.
When the LOS link is guaranteed, FSO-based backhaul can offer several Gbps data rates with a much lower deployment cost than the optical fiber-based backhaul.
When the LOS link is absent, optical IRS, a planar metasurface that controls the optical propagation environment by adjusting the phase shifts of reflecting elements can be used to offer the virtual LOS link between the BSs and CU. 

\smallskip \noindent
\textbf{Rank-sufficient THz LOS communications using cell-free massive MIMO network:} 
In the THz band, due to the severe path loss and high directivity, the conventional MIMO techniques exploiting the rich scattering of multipath environments will not work.
In the case where the LOS path is the dominant, the cell-free massive MIMO systems can be an appealing solution to address the rank-deficiency issue of THz communications. 
Specifically, through the joint transmission of small BSs, the cell-free massive MIMO systems can generate a virtual multipath propagation. 
Further, the shortened communication distance along with the extremely large number of transmit antennas incur a spherical wavefront which creates a rich non-linear pattern of phase variations and therefore, increases the channel rank. 

\subsection{Boundless and Fully Integrated Terrestrial $\&$ Non-Terrestrial  Networks}

In remote and rural areas where the existing terrestrial cellular network is unreachable \cite{6GNTN}, NTNs will be a useful means of providing cost-effective and high-speed internet services. 
Depending on the flying altitude and coverage area, NTN devices can be classified as UAV, HAPS, and satellite.
Satellite stations are further divided into geostationary orbit  (GEO),  medium  earth  orbit  (MEO),  and low  earth  orbit (LEO) based on their orbital altitude.

In Fig. \ref{fig:3D_connectivity}, we plot the round-trip delay (RTD) and single satellite coverage on the Earth's surface for various altitude levels.
When a satellite revolves in orbit close to the surface of the earth, a faster response time can be guaranteed, but the satellite's decreases owing to its proximity to the ground.
%
%
For example, only three equally spaced GEO satellites are needed to can provide near-global coverage because each GEO satellite covers almost 35\% of the Earth's area, but RTD in this case exceeds 0.5 second. Conversely, the RTD of the LEO satellite is less than 0.1 second, but the coverage will be just 0.2\% at an altitude of 550 km. 
One can infer from this discussion that a large number of LEO satellites, often called LEO mega-constellation, would be a good option to guarantee the latency in the order of a few tens of milliseconds and global coverage.
%

We note that there exists a trade-off between the ground observer's minimum elevation angle and link quality (see Fig. \ref{fig:3D_connectivity}). 
If we increase the minimum elevation angle, so is the LOS probability of the satellite link; however, the beam coverage and visible time of the satellite will be reduced.
%
To balance this trade-off, the ground station’s minimum elevation angle should be carefully determined in practice.
%
%
%
Recently, SpaceX decided to change the minimum elevation angle of Starlink LEO satellites from 40\textdegree\ to 25\textdegree\ (granted by the FCC in April 2021) to improve coverage with a small number of LEO satellites \cite{starlink}.

In analyzing the NTN communication of satellite and aerial devices, areal capacity ($\mathrm{bits/s/m^2}$) is no longer useful; therefore, a new metric called volumetric traffic capacity ($\mathrm{bits/s/m^3}$) that measures the quality of user experience in 3D space is needed.
%
%
%
To obtain a sufficient volumetric traffic capacity to ensure universal global connectivity for 6G systems, the following issues must be addressed.

\smallskip \noindent
\textbf{Mobility-aware channel modeling and tracking in 3D space:} Since the NTN terminals are located at a high altitude and are moving fast with respect to the ground stations, LOS propagation paths are guaranteed, but the wireless signals suffer significant Doppler shift. The first step in addressing the distortions caused by high mobility and a large coverage area is an accurate 3D channel model that accounts for the Doppler shift and propagation delay.
In addition, dynamic channel prediction based on the real-time location tracking of NTN terminals is of great importance 
to achieve agile and effective NTN management. 


\smallskip \noindent
\textbf{Boundless multi-layered TN/NTN architecture:}  To provide sufficient and continuous coverage everywhere, the large LEO satellite constellations should be combined with GEO/MEO using the coverage-aware protocol and NTN-backhaul connection. In fact, a boundless multi-layered architecture fully integrating TNs and NTNs is essential to guarantee the reliable connection in the ocean and desert, as well as high-speed trains and airplanes \cite{38821}. UAV and HAPS play a key role in improving the connection quality because they can be used as relays between the LEO/MEO/GEO satellite and ground stations. In addition, an integrated network protocol that includes multi-layer routing, network mobility management, and resource allocation should be introduced.

\smallskip \noindent
\textbf{Seamless inter-beam/satellite/orbit handover:}
    Owing to the high-speed movement of the LEO satellites, the maximum visible time of the satellites at the ground station is at most 10$\sim$20 minutes, causing a frequent handover even for a fixed ground station. 
    To mitigate this problem, LEO satellites should be connected to nearby satellites in the same or adjacent orbits through laser or radio frequency (RF) links. This so-called inter-satellite link (ISL) connectivity supports relatively low end-to-end latency even for the long distance communication scenario.
   Candidate techniques to establish an efficient ISL connectivity include spatial-mode diversity and multiplexing, wavelength division multiplexing, and adaptive beam control using pointing/acquisition/tracking (PAT) process.
    %
    Another challenge of NTNs is that the cell size varies according to the elevation angle of the NTN terminals; therefore, conventional handover techniques might not be suitable.
    %
    %
    To address this challenge, a forecast-based handover decision on the basis of satellite position estimation is required.
    %
    %

\begin{figure}[!t]
\centering
\includegraphics[width=3.7in]{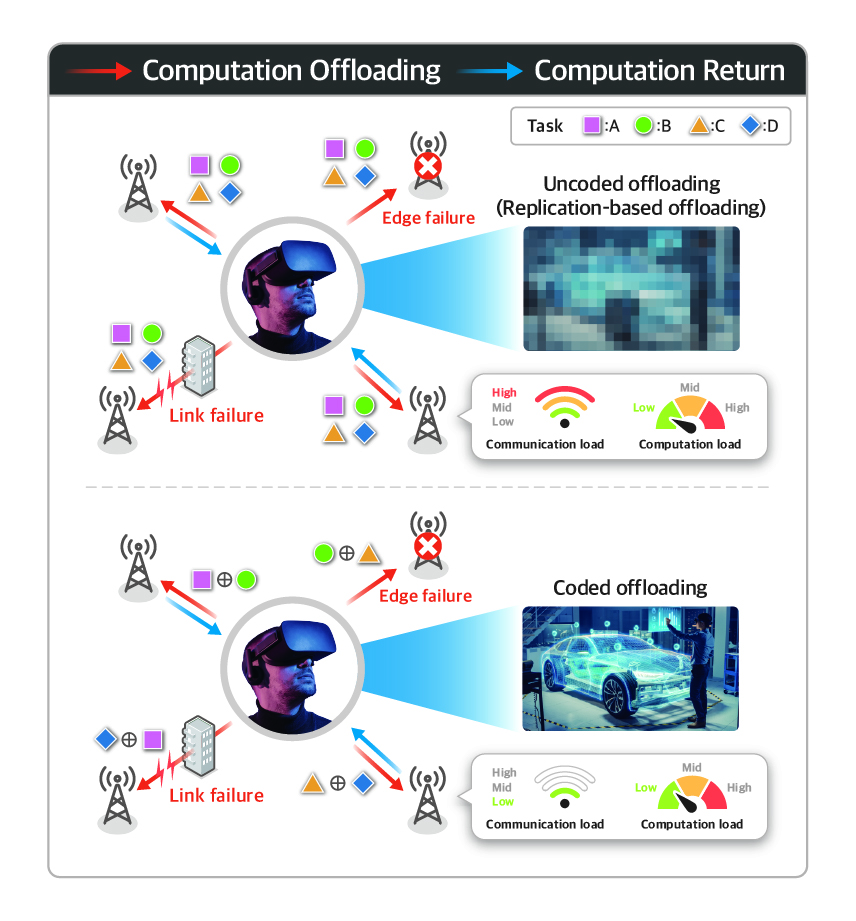} 
\caption{Over-the-air distributed computing scenario where a VR headset assigns linear computation tasks to four edges with uncoded/coded offloading.}
\label{fig_comp}
\end{figure}

\smallskip
\noindent
\textbf{3D cell planning and fast beam management:}
Since the cell size varies according to the altitude and elevation angle of the satellite, we need to introduce 3D cell planning based on the orbits of satellites. 
A satellite with a higher altitude and lower elevation angle tends to form a larger cell size owing to its wider beam footprint on the ground. 
Specifically, the GEO and LEO satellites have beam widths of up to 3,500 km and 1,000 km at an altitude of approximately 600 km, respectively. 
%
%
%
%
In addition, 3D cell types can be classified into earth-moving and earth-fixed cells with respect to the Earth's surface, depending on whether the satellite beams can be fixed or steered. 
The earth-fixed cell should be designed to be highly robust to satellite beam pointing errors since even a very small error makes the satellite beam footprint placed tens of kilometers away from the initially designated area on the Earth; whereas, the earth-moving cell may require more frequent handovers because of the dynamics of beam movement on the ground. 
%
As for the cell planning in multi-layered TN/NTN architecture, a truncated octahedron-shaped cell planning distinguished from the hexagon-shaped 2D cell layout is desired when considering the volumetric quotient in 3D space. 


\smallskip
\noindent
\textbf{Intelligent spectrum management and NTN system optimization:}
To achieve seamless and broad coverage in integrated terrestrial and non-terrestrial networks, powerful spectrum management scheme is needed.
%
To this end, intelligent and dynamic cooperation between primary and cognitive users is preferred over the semi-static spectrum sharing.
%
%
Note that because the S-band (i.e., 2170$\sim$2200 MHz) is adjacent to the 3G/4G LTE band (i.e., 2110$\sim$2170 MHz), the large Doppler shift of the high-speed satellites causes severe interference in the adjacent band, deteriorating the service quality of 3G/4G TN services.
A possible option to mitigate the interference is the on-board pre-compensation  of the Doppler shift  at the center of the beam on the ground while the residual Doppler shift can be corrected at each receiver side. 
To further improve spectral efficiency, we should have an appropriate frequency reuse pattern among multiple beams, available channels, and antenna gains.

\begin{figure}[!t]
\centering
\includegraphics[width=3.2in]{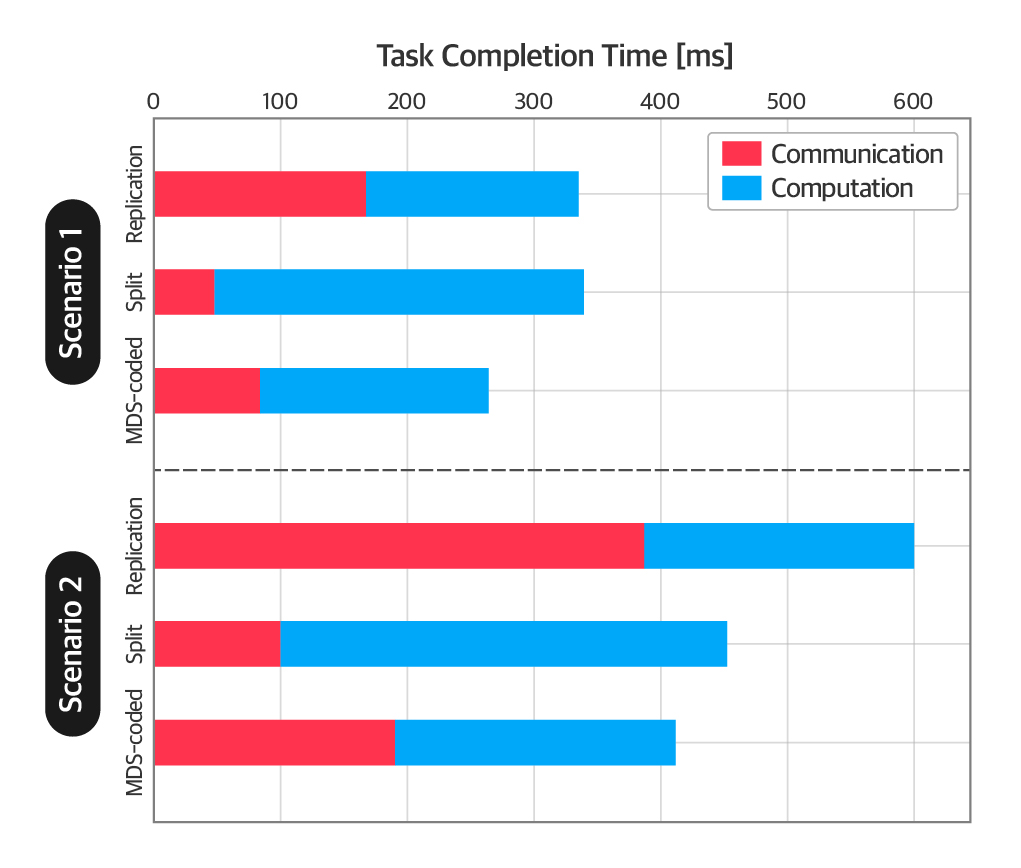} 
\caption{Task completion times when a user offloads linear computation tasks to four edges over two different channel conditions (Scenario 1: LOS channel at a distance of 15m; Scenario 2: NLOS channel at a distance of 300m).}
\label{fig_comp2}
\end{figure}

\begin{figure*}[!t]
\centering
\includegraphics[width=6.5in]{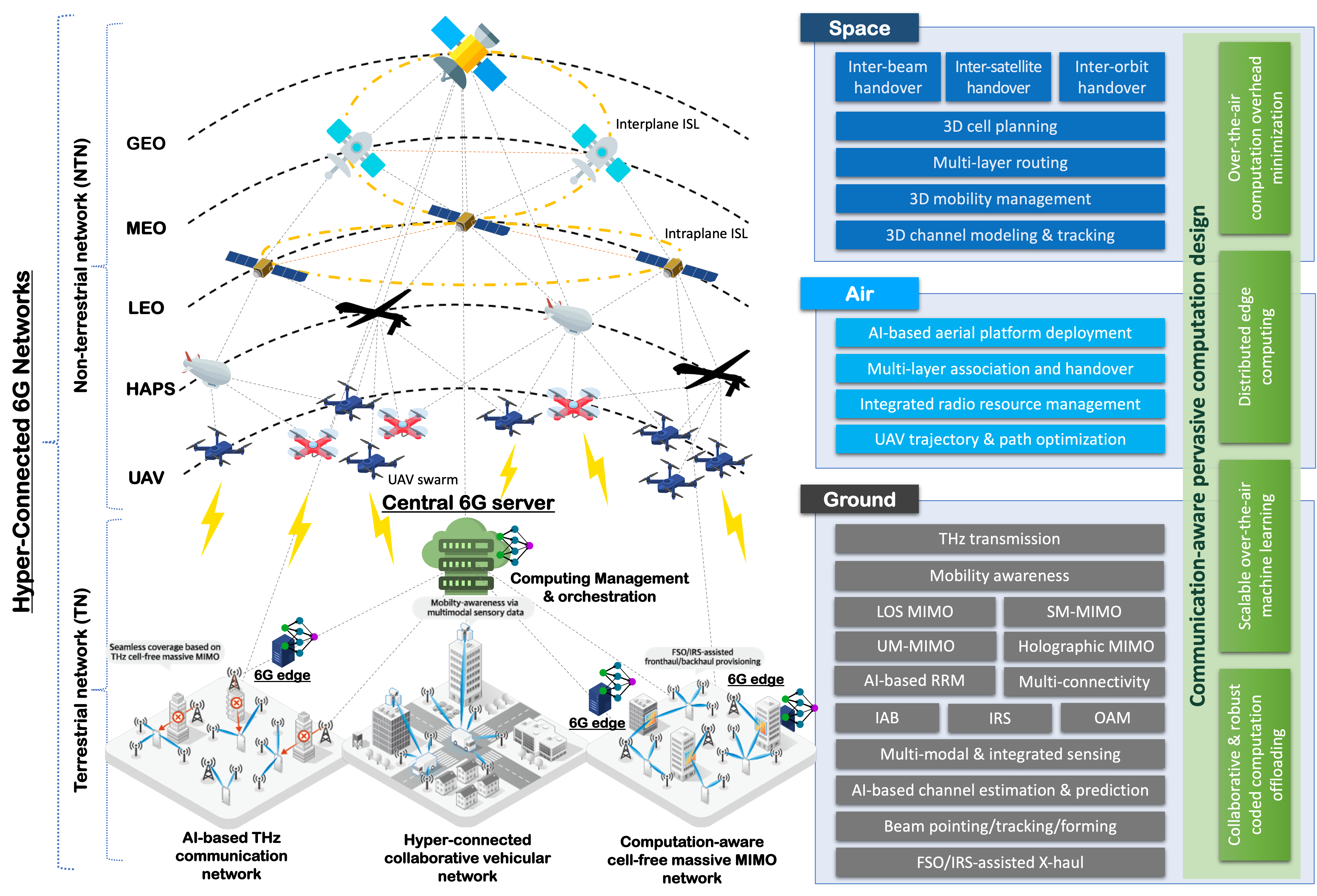} 
\caption{Hyper-connected 6G network architecture}
\label{fig_NWA}
\end{figure*}

\subsection{
Over-the-air Distributed Computing for Artificial Intelligence Applications}

A future trend in dealing with the relentless growth in computing-intensive applications is pervasive computing, in which mobile devices offload computational overhead to network edges or nearby devices. 
Notwithstanding the rosy outlook, a number of issues have to be addressed since there is a fundamental trade-off between computation and communication. In fact, when one tries to offload computational overhead, one should investigate the overhead of the communication link.
%
Fig. \ref{fig_comp} depicts computation offloading scenarios where a VR headset assigns linear computations to four edges with various offloading schemes \cite{CodedComp}.
%
Since the reduction in the communication load over the network increases the computation overhead at the edges, we need to come up with a proper offloading scheme optimized for the computation capability and network conditions.
%
%
%
One thing to note is that the wireless channels experienced by devices and their computing capabilities are quite different among devices so that there is a large variation in the task completion time.
Particularly, this problem is critical for the ultra-reliable and low-latency communications (URLLC) services since the task completion time is determined by the device with the worst channel condition or straggling/malfunctioning devices (e.g., edge failure in Fig. \ref{fig_comp}).
%
%
%
For the seamless integration of communication and computing,
the following approaches can be considered.

\smallskip \noindent
\textbf{Communication-aware pervasive computation design: } To alleviate the difficulties caused by the channel uncertainty, we need to exploit the mmWave and THz frequency bands.
    If sufficient bandwidth is provided and LOS channel conditions are guaranteed for every computing node, wireless links will not be a serious bottleneck.
    Obviously, this ideal condition cannot be guaranteed always, so we need to come up with an environment-aware and communication-friendly computing mechanism.
    %
    For example, the communication load can be reduced by quantization, compression, and sparsification of the transmit information as long as it does not severely degrade the task quality.
    For special cases where the (weighted) average of the computation tasks is needed, we can design the system such that the computation results are added during the wireless transmission of the same time/frequency resources. This so-called over-the-air computation can significantly reduce the communication and computation overhead.

\smallskip \noindent
\textbf{Collaborative and robust coded computation offloading:} A naïve approach to exploit multiple computing nodes is to split the whole computation task into multiple pieces and then distribute them equally to multiple nodes. This split-based offloading will not be effective in the presence of link failure and straggling/malfunctioning devices, since the whole task can be 
    completed only when all subtasks are finished. One way to avoid this problem is to replicate a part of computation tasks and assign them to the multiple devices. 
    %
    
One option to overcome difficulties caused by the channel heterogeneity, link failure, and straggling/malfunctioning is to apply coding techniques to distributed task allocation. 
    By introducing redundancy in distributed tasks, coded distributed computing can be more resilient to the link and edge failures. For example, if computation tasks are designed using maximum distance separable (MDS) codes, one can recover the straggling tasks from other completed 
    tasks \cite{CodedComp2}.
    As illustrated in Fig. \ref{fig_comp2}, one can achieve the better computation-communication trade-off by assigning coded computation sub-tasks to computing nodes.

\smallskip \noindent
\textbf{Scalable and over-the-air deep learning:} 
To perform a large-scale deep learning (DL) task over wireless networks, we must decide how to allocate the tasks among the cloud, network edges, and the mobile devices \cite{scalable}.
For instance, network edges perform DNN training, and the mobile device performs the inference using the trained model from edges.
To perform the flexible DL using abundant network resources, we can consider the following: 1) federated learning, which trains local models on each device and then passes updated parameters to the central cloud for a global model; 2) distributed learning, in which each device communicates directly with its neighbors; and 3) split learning, in which devices or the central cloud trains each split model individually.

These DL techniques can be combined with over-the-air techniques \cite{air} to effectively aggregate large-scale distributed data in ultra-low-latency
scenarios. In addition, a scalable approach can be applied to resource-constrained mobile devices to reduce communication costs by leveraging gradient clustering, correction, and quantization methods.

\section{Summary and Outlook}

In this article, we discussed the vision and challenges to achieve the hyper-connected society in 6G.
We have identified several key enabling technologies and their roles in supporting ultra-broadband, ubiquitous 3D, and computation-aware connectivity in 6G. With these disruptive technologies, 6G is envisioned to collapse the existing boundaries of space/air/ground networks, communication/computing/intelligence capabilities, as well as physical/virtual worlds. 
We anticipate that this will bring the truly {\it 6G hyper-connectivity} society to humans, machines, and sensors in ubiquitous 3D global coverage (beyond ground-level 2D connectivity for 5G) even in space, oceans and desserts. 
In Fig. 6, we summarized the hyper-connected 6G network architecture by highlighting the key enabling technologies along with other remaining major challenges.

Going forward, achieving Tbps data rates, zero coverage holes, and pervasive computing for connected intelligence
will contribute to reducing differences in social and regional infrastructures and economic opportunities, thereby ameliorating many social issues such as the digital divide, regional polarization, and education inequality.
Our hope is that this article will spark further interest and expedite the research activities for 6G.

\ifCLASSOPTIONcaptionsoff
  \newpage
\fi

\end{document}